\documentclass[lettersize,journal]{IEEEtran}

\usepackage[ left=0.625in, top=0.75in, right=0.75in, bottom=1.25in, headsep=0.25in, headheight=20pt, heightrounded, letterpaper] {geometry}
\usepackage[utf8]{inputenc}
\usepackage{lipsum}
\usepackage{comment}
\usepackage{xfrac}  

\usepackage{lipsum,microtype}
\usepackage{authblk}
\usepackage[ruled,vlined]{algorithm2e}
\usepackage{amsmath, nccmath}
\usepackage{geometry}
\usepackage{tabularx}
\usepackage{booktabs}
\usepackage{setspace}
\usepackage[labelfont=bf,format=plain,justification=raggedright,singlelinecheck=true]{caption}
\usepackage{mathtools}
\usepackage{amssymb}
\usepackage{caption}
\usepackage{xcolor}
\usepackage{subcaption}
\usepackage{graphicx}
\graphicspath{{./Fig/}} 

\usepackage[ruled,vlined]{algorithm2e}
\usepackage{cite}
\newcounter{mycounter}
\usepackage{subfig} 
\usepackage{fancyhdr}
\usepackage{textcomp}
\usepackage{blindtext}
\usepackage{optidef}
\usepackage{subcaption} 
\usepackage{ragged2e} 
\usepackage{multirow}
\usepackage{longtable}
\usepackage{pifont}
\definecolor{orangish}{wave}{620}

\usepackage{fancyhdr}
\usepackage{textcomp}
\usepackage{blindtext}
\title{Prioritized Value-Decomposition Network for Explainable AI-Enabled  Network Slicing}

\author[1]{Shavbo Salehi}
\author[2]{Pedro Enrique Iturria-Rivera}
\author[2]{Medhat Elsayed}
\author[2]{\\Majid Bavand}
\author[3]{Raimundas Gaigalas}
\author[2]{Yigit Ozcan}
\author[1]{Melike Erol-Kantarci, Senior Member, IEEE}
\affil[1]{School of Electrical Engineering and Computer Science, University of Ottawa, Ottawa, Canada}
\affil[2]{Ericsson Canada Inc., Ottawa, Canada}
\affil[3]{Ericsson AB, Stockhom, Sweden}
\affil[ ]{{Emails: \{ssale038, pitur008, melike.erolkantarci}\}@uottawa.ca}
\affil[ ]{{\{medhat.elsayed,majid.bavand, raimundas.gaigalas,yigit.ozcan}\}@ericsson.com}

\date{}
\begin{document}

\twocolumn

\maketitle
\begin{abstract}

Network slicing aims to enhance flexibility and efficiency in next-generation wireless networks by allocating the right resources to meet the diverse requirements of various applications. Managing these slices with  machine learning (ML) algorithms has emerged as a promising approach however explainability has been a challenge. To this end, several Explainable Artificial Intelligence (XAI) frameworks have been proposed to address the opacity in decision-making in many ML methods. In this paper, we propose a Prioritized Value-Decomposition Network (PVDN) as an XAI-driven approach for resource allocation in a multi-agent network slicing system. The PVDN method decomposes the global value function into individual contributions and prioritizes slice outputs, providing an explanation of how resource allocation decisions impact system performance. By incorporating XAI, PVDN offers valuable insights into the decision-making process, enabling network operators to better understand, trust, and optimize slice management strategies. Through simulations, we demonstrate the effectiveness of the PVDN approach with improving the throughput by 67\% and 16\%, while reducing latency by 35\% and 22\%, compared to independent and VDN-based resource allocation methods. 

\end{abstract}


\section{Introduction}
\label{section:intro}

In a multi-agent network slicing system, the challenge lies in meeting distinct quality of service (QoS) requirements, such as ultra-reliable low latency communications (URLLC) demanding low latency for real-time applications, while enhanced mobile broadband (eMBB) prioritizing high throughput for data-intensive services \cite{liu2023network}. These various requirements highlight the need for intelligent resource allocation strategies that can dynamically adapt to changing network conditions and user demands \cite{cai2023deep}. However, the complexity of the underlying algorithms, often driven by ML models, can make it difficult for network operators to understand and trust the decision-making processes \cite{donatti2023survey}. One of the solutions that have been highlighted is the benefits of explainability \cite{singh2023directive}. By providing insight into ML-based decisions, explainability tools help operators understand the reasons for allocating resources to one slice over another, ensuring that the system functions as intended. These tools also help diagnosing issues more effectively and making informed adjustments to optimize network performance \cite{fiandrino2023explora}.

Explainability involves using model-agnostic methods to describe or clarify the model's internal functions \cite{brik2023survey}. Explainability can be categorized into intrinsic and post-hoc methods. 
Intrinsic methods build inherently interpretable policies during training, whereas post-hoc methods provide explanations after a policy has been trained \cite{boggess2023explainable}. Understanding and managing the behavior of multi-agent systems, particularly in complex scenarios such as joint resource allocation in network slicing, can greatly benefit from these explainability methods.

One of the key challenges in joint resource allocation within a multi-agent system is managing conflicting agent priorities while ensuring efficient, fair resource utilization in network slicing. 
To this end, using the value-decomposition network (VDN) as an explainable method involves breaking down a complex joint value function into simpler, agent-specific value functions \cite{VDN}. This decomposition can help in understanding and explaining the contributions of URLLC and eMBB slice agents to the overall system's decisions. In this paper, we focus on post-hoc methods for resource allocation in a multi-agent network slicing scenario, aiming to improve resource allocation to meet the requirements of the URLLC and eMBB slices for a multi-agent system by a prioritized VDN (PVDN). This method by prioritizing the resource allocation to a slice chooses better action and then using the knowledge of this for another slice, to improve the resource allocation in the system. 
This approach prioritizes resource allocation for a single slice, determining the optimal action based on its specific needs, and then applies the insights gained from this process to improve the efficiency and effectiveness of resource allocation for other slices, ensuring that the system adapts dynamically to the varying requirements of different network slices. 

\section{Literature Review}
\label{section:LR}
Centralized approaches have been used in multi-agent systems in some works, but such methods were found to fail in cooperative multi-agent reinforcement learning (MARL) \cite{VDN}. A decomposed reward function (DRF) in reinforcement learning (RL) refers to the breakdown of a complex reward function into simpler and more manageable components. Each component corresponds to specific aspects of the task or the environment, making it easier to understand and optimize the overall objective. DRF is particularly beneficial in multi-agent systems, where different agents have distinct objectives. In \cite{juozapaitis2019explainable}, the DRF of Markov decision processes (MDP) is introduced which shows how much an agent prefers one action over the other by defining a vector-valued reward function. 

Solving the problem of network slicing using an MDP is challenging due to the complexity of joint action-value functions in MARL systems. Each agent must consider the actions of another agent, leading to a combination of explosion of potential joint actions. This makes the learning process inefficient and computationally intensive. Moreover, coordinating these agents to achieve optimal network performance adds another layer of difficulty. VDN addresses these challenges by decomposing the joint action-value function into individual value functions for each agent \cite{VDN}, addresses cooperative MARL challenges by introducing a VDN architecture that decomposes the team value function into individual agent-wise value functions, and demonstrating superior performance in partially-observable multi-agent domains. Furthermore, DRFs enhance explainability in multi-agent systems by breaking down complex objectives into individual components. 
In \cite{singh2024network}, extended t-statistic feature selection as an explainable ML technique is proposed for network slicing in multi-class classification.
In \cite{ameur2024leveraging}, a composable explainable RL (XRL) framework for 6G network slicing, using large language models (LLMs) and prompt engineering is proposed to enhance the interpretability of deep RL (DRL) decisions. In \cite{fiandrino2023explora} EXPLORA, a framework that provides explainability for DRL in O-RAN is proposed by linking DRL actions with input states using an attributed graph, thereby improving transmission bitrate and tail. In \cite{barnard2022resource}, the application of explainable AI (XAI) techniques to short-term resource reservation is explored, demonstrating kernel Shapley additive explanation effects on clarifying AI model decisions and aiding in fault diagnosis. 

In this paper, we propose PVDN, which incorporates VDN while accounting for priority in the system to improve the overall reward in a multi-agent environment. Unlike traditional transfer learning, where one agent is trained and its knowledge is transferred to another via a mapping function \cite{Globecom2023}, PVDN allows both agents, each with distinct objectives, to train simultaneously. Moreover, PVDN outperforms a centralized approach, self-play ensemble Q-learning, by prioritizing agents based on the specific requirements of URLLC and eMBB slices, thus providing better coordination between latency-sensitive and throughput-oriented services \cite{salehi2024self}. Furthermore, PVDN differs from hierarchical Q-learning \cite{pateria2021hierarchical}, which prioritizes important actions and allocates resources based on initial requirements. In PVDN, actions are not prioritized in advance, but subtask priorities are respected during learning without any trade-off between conflicting tasks. This approach offers a clear framework for addressing complex RL problems and provides deeper insights into subtask composition. PVDN aims to reuse and adapt subtask solutions from a slice that was learned previously to address the requirements of a new slice. Therefore, PVDN requires a decomposed learning algorithm that models individual subtask solutions and combines them to form a solution for the network slicing scenario.

\section{System Scenario}
\label{section:sce}

In this paper, a multi-agent scenario for eMBB and URLLC slices is assumed in which two agents, eMBB slice management agent (MSMA) and URLLC slice management agent (USMA) assign available resources to various user equipment (UEs). 
The problem formulation for the MSMA is described in eq. (\ref{eqn:rewardMSMA}). The MSMA aims to increase the system's throughput, while the USMA focuses on reducing system latency. 

\begin{equation}
\begin{aligned}
& \max \quad B^{eMBB,avg}_{MSMA}\\
\end{aligned}
\label{eqn:rewardMSMA}
\end{equation}

\begin{equation}
\setcounter{mycounter}{1}
\tag{\themycounter.a}
\sum_{u\in M^{eMBB}_{MSMA}} x_{u,r'} =1  
\label{eqn:1a}
\end{equation}

\begin{equation}
\setcounter{mycounter}{1}
\tag{\themycounter.b}
\sum_{r' \in N }( \sum_{u\in M^{eMBB}_{MSMA}} x_{u,r'}) \leq |N|
\label{eqn:1b}
\end{equation}

\begin{equation}
\setcounter{mycounter}{1}
\tag{\themycounter.c}
 C^{eMBB}_{MSMA} \leq C
 \label{eqn:1c}
\end{equation}
where $B^{eMBB,avg}_{MSMA} = \frac{\sum_{u \in M^{eMBB}_{MSMA}} B^{eMBB}_{MSMA,u}}{|B^{eMBB}_{MSMA}|}$ shows the average throughput in the eMBB slice for all $u \in UEs$ that the MSMA aims to boost. The MSMA endeavors to assign each RB to a UE as depicted in eq. (\ref{eqn:1a}), in which $x_{u,r'}$ is a binary variable indicating that resource block $r$ is assigned to UE $u$ in eMBB slice. It should be noted that eq. (\ref{eqn:1b}) ensures that the total allocated resource blocks equal the number of available resources. Finally, eq. (\ref{eqn:1c}) denotes that the assigned computation capacity should remain within the limits of the available resources. $B^{eMBB}_{MSMA}$ is related to the downlink capacity between UE $u$ and the eNB, which is calculated by:

\begin{multline}
    B^{eMBB}_{MSMA} = \displaystyle\sum_{r \in N_u} b_{RB}log \Bigr(1+ \\ \frac{p_{j,r}x_{j,u,r}q_{j,u,r}}{b^{RB}N_{0}+\Sigma_{j' \in J_{-j}}\Sigma_{u' \in u_{j'}}\Sigma_{r' \in N_{j'}}p_{j',r'} X_{j',u',r'} g_{j',u',r'}}\Bigr),\\
\label{eqn:eq2}
\end{multline}
where $N_u$ is the set of allocated RBs to the UE $u$, $b_{RB}$ is a RB's bandwidth, $N_{0}$ is noise power density, $p_{j,r}$ is the transmission power of RB $r$ in the eNB $j$. $x_{j,u,r}$ indicates a binary variable that illustrates whether RB $r$ in eNB $j$ is assigned to the UE $u$ or not. $q_{j,u,r}$ refers to the channel gain of the eNB $j$ and UE $u$. $J'$ indicates the set of eNBs, except the target eNB $j$, $u'$ is the set of UEs in other eNBs, and $N'$ is the set of total RBs in other eNBs. The problem formulation for USMA can be found in eq. (\ref{eqn:rewardUSMA}).

\begin{equation}
\begin{aligned}
& \max \quad D^{tar}_{USMA}-D^{URLLC,avg}_{USMA}\\
\end{aligned}
\label{eqn:rewardUSMA}
\end{equation}

\begin{equation}
\setcounter{mycounter}{3}
\tag{\themycounter.a}
\sum_{v\in M^{URLLC}_{USMA}} x_{v,r'} =1  
 \label{eqn:2a}
\end{equation}

\begin{equation}
\setcounter{mycounter}{3}
\tag{\themycounter.b}
\sum_{r' \in N_{j'}} ( \sum_{v\in M^{URLLC}_{USMA}} x_{j,v,r'}) \leq |N_{j}| 
 \label{eqn:2b}
\end{equation}

\begin{equation}
\setcounter{mycounter}{3}
\tag{\themycounter.c}
 C^{URLLC}_{j} \leq C_{j}
  \label{eqn:2c}
\end{equation}
where in eq. (\ref{eqn:rewardUSMA}), $D^{URLLC,avg}_{USMA} = \frac{\sum_{v \in M^{URLLC}_{USMA,v}} D^{URLLC}_{USMA}}{|D^{URLLC}_{USMA}|}$ shows the average latency in the URLLC slice in which the USMA tries to reduce, and $D^{tar}_{USMA}$ the target delay of the URLLC slice. $M^{URLLC}_{USMA}$ is an indicator of how many URLLC slices are available. Using eq. (\ref{eqn:2a}) ensures that each RB is allocated to each $v \in$ UE of the URLLC queue. According to eq. (\ref{eqn:2b}) and eq. (\ref{eqn:2c}), the total number of RBs and computation capacity allocated to the eNB should not be greater than the number of resources available within the eNB.

To evaluate the network delay, we consider the effect of three criteria, including transmission delay, re-transmission delay, and queueing delay which are calculated by $D^{URLLC,avg}_{USMA} = d^{Tx}+d^{rTx}+d^{que}$. $d^{edge}$ illustrates the queueing delay for completing a task that is calculated by the number of central processing unit (CPU) cycles, as $d^{edge} = \frac{c_{u,q}}{\beta C}$, where $c_{u,q}$ represents required computation resources of the UE $u$ for task $q$, $\beta \in [0,1]$ refers to the computation resources assigned to a task, and $C$ indicates the total MEC server's computation capacity. In this scenario, the radio resources refer to time-frequency resource blocks and to ensure meeting both slices' requirements,  it is critical to ensure that resources are allocated in an efficient and timely manner. 

\section{Proposed Method}
\label{section:methods}

We consider a multi-agent system with two agents, MSMA and USMA, for managing eMBB and URLLC slices, respectively. The agents execute actions at $t = 1, 2, 3, ...$ receiving observations and rewards for maximizing cumulative rewards. The problem is formulated as an MDP problem by tuples $<s,a,\mathcal{T},r>$, characterized by the following components:

\begin{itemize}
    \item State $(s_t)$: The state represents the current UEs tasks at time $t$ in the queue for both slices, illustrating by $s_t^{i}$, $i \in {MSMA,USMA}$.

    \item Action $(a_t)$: Each agent allocates radio resource blocks to its respective slice, denoted by $a_t^{i}$.

    \item Transition Function $(\mathcal{T})$: The transition function $T(s^i_{t+1}|s^i_t,a^i_t)$, defines the probability of moving to the next state $s_{t+1}$ given the current state $s_t$ and action $a_t$. This function, $\mathcal{T} : s_t \times a_t \rightarrow P(s_t)$ captures the network dynamics and how resource allocation decisions impact future network conditions.

    \item Reward $(r_t)$: The reward function $r_t(s_t,a_t)$ is designed to capture the overall network performance. 

\end{itemize}
MSMA and USMA aim to maximize the expected cumulative discounted reward, $r_t = \sum_{t=1}^{TTI} \gamma_{t-1} r_t$, with $\gamma$ as the discount factor. Their actions are selected based on a policy function, $\pi: s_t \rightarrow P(a_t)$, where $P$ is the probability distribution over the action set. Solving network slicing with MDP and Q-learning is challenging due to the complexity of joint action-value functions in multi-agent systems, leading to inefficiency and computational intensity in coordinating agents for optimal performance. VDN addresses these challenges by decomposing the joint action-value function into individual value functions for each agent. This simplification allows each MSMA and USMA to independently learn its optimal policy while still contributing to the overall system's performance. 

\subsection{Value Decomposition Reward}
\label{subsection:VDN}

In fully observed environments, agents have access to stationary optimal policies. However, in the MARL system, each agent—MSMA and USMA—has access only to its local observations, resulting in a partially observed environment. Observations and actions are distributed across both agents, which complicates the process of predicting outcomes, as each agent operates with incomplete information. To address this challenge, the policy incorporates past agent observations and actions from the history, as shown in the equation below:

\begin{equation}
    h_t^i = (a_1^i o_1^i r_1^i, \dots, a_{t-1}^i o_{t-1}^i r_{t-1}^i), \quad \text{for} \, i \in \{\text{MSMA}, \text{USMA}\}
\end{equation}
Since the environment is partially observable, the Q-function is updated using the history, $Q^i(h^i_t,a^i_t)$, to account for incomplete information, in which $h_t = (h_t^{MSMA}, h_t^{USMA})$ represent the tuple of MSMA and USMA histories. Then joint policy is a map of $\pi: h^{i} \rightarrow p(o^{i}),  i \in \{\text{MSMA}, \text{USMA}\}$ and the policies for each agent are independently determined. In this system, MSMA and USMA are responsible for distinct slices, with their individual observation sets $o^{eMBB}$ and $o^{URLLC}$, and actions $a^{eMBB}$ and $a^{URLLC}$. Each agent receives only the joint reward, and the final objective is to optimize the total reward $r_t$:
\begin{equation} 
r_t = \gamma_{t-1} r_{t-1}^{i} \quad \text{where } i \in { \text{MSMA, USMA} } 
\end{equation}
The value function is then calculated as follows:
\begin{multline}
    V^{\pi}(s) := \sum_{i\in{USMA,MSMA}} \\
    \mathbb{E}\left[\sum_{t=1}^{TTI} \gamma^{t-1}r^i(s^i_t,a^i_t,s^i_{t+1}) \mid s^i_1=s^i; a^i_t \sim \pi(0|s^i_t)\right]
\end{multline}
where, each agent’s value function reflects the expected reward based on its actions, with the total reward having an additive nature. To accelerate learning, a dueling architecture is employed, representing the Q-value as a combination of a value and an advantage function. Multi-step updates with forward view eligibility traces are used over a certain number of steps. The action-value function is then represented by:
\begin{multline}
    Q^{\pi}(s^i, a^i) := \mathbb{E}_{s^{i'} \sim \mathcal{T}(. | s^i, a^i)} \left[ r^i(s'^{i}, a^i, s^i) + \gamma V(s'^{i}) \right], \\
    \quad i \in \{MSMA, USMA\}
\end{multline}
The optimal value function $V^*(s^i)$ and policy are computed as $V^*(s^i) = \sup_{\pi} V^{\pi}(s^i)$ and $Q^*(s^i) = \sup_{\pi} Q^{\pi}(s^i, a^i)$. For this scenario, the greedy policy is calculated by:
\[
\pi(s^i) := \arg\max_{a^i \in A} Q(s^i,a^i), \quad i \in \{MSMA, USMA\}
\]
VDN
effectively decomposes the joint value function into individual agent value functions, facilitating learning and deployment in multi-agent systems. However, its main shortcoming lies in the inability to account for the varying importance of different tasks, as it treats all agents equally in the resource allocation process; PVDN addresses this limitation by prioritizing tasks and ensuring that critical agents receive the appropriate focus, ultimately optimizing the system's overall resource allocation.

\subsection{Prioritized Value Decomposition Network}
\label{subsection:PVDN}

For prioritizing agents in the system, we consider a prioritized version of the VDN. In the PVDN, the joint action-value function is factorized into individual value functions for each agent. This decomposition simplifies the learning process, enabling MSMA and USMA to learn their policies more effectively by focusing on local observations. 
Then PVDN learns the optimal linear value decomposition from the joint reward signal, where $Q^{i}$ represents the individual value functions. This method helps mitigate issues related to spurious reward signals that arise from independent learning. Although agents learn in a centralized manner during training, they can operate independently during deployment. The assumption here is that the joint action-value function can be additively decomposed, as shown in the equation below:
\begin{equation} 
Q((h^{i}), (a^{i})) \approx \ \sum_{i \in {\text{MSMA}, \text{USMA}}} Q^i(h^i, a^i) 
\end{equation}
then by combining the Q-functions, PVDN 
selects the maximizing action for its policy. Implementing PVDN effectively enforces task priorities, ensuring that low-priority subtasks with significant reward scales do not dominate high-priority subtasks. Therefore, Q-decomposition requires transforming subtasks to allow for the summation of rewards and Q-functions.

PVDN approach aims to prioritize agents based on their roles in the reward function. While the USMA has priority in choosing actions compared to the MSMA, changes in eMBB UEs’ requirements or an increase in requests might lead to meeting the eMBB slice requirements having a greater effect on the overall reward function. To address this, we implement a weighted reward function where the system's weights dynamically adjust according to agents in both slices of the system, which are affected by changes in the number of UEs, their requirements, and the agents action selection.
\begin{figure}
    \centering
    \includegraphics[width=0.85\linewidth]{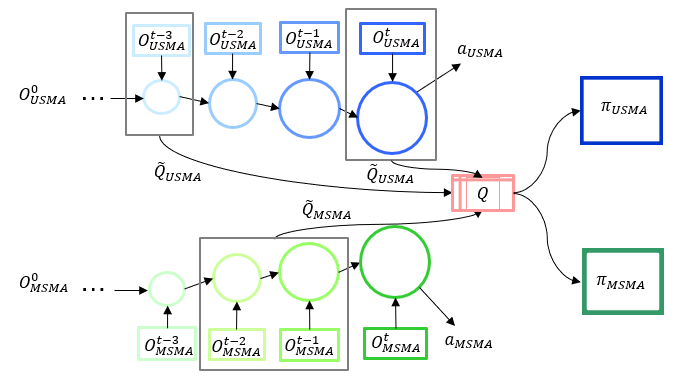}
    \caption{Prioritized Value Decomposition Network}
    \label{fig:PVDN}
\end{figure}
It should be noted that in the considered MARL system, the complex task’s reward function is the sum of sub-task rewards, $r(s_t,a_t) = \sum_{i \in \{\text{MSMA}, \text{USMA}\}} r^i(s^i_t, a^i_t)$. 

By the reward shaping, the rewards are given to each agent based on their cooperation. For example, if URLLC takes an action that improves latency but slightly decreases throughput, the system might still reward eMBB so that it doesn't discourage cooperative behavior. In the PVDN, the reward for encouraging agents to have more cooperation is modified as below:

\begin{multline}
    r = \omega_{UMSA} \cdot (r_{USMA}-\beta \cdot \Delta B^{eMBB,avg}_{MSMA}) \\
    + \omega_{MSMA} \cdot (r_{MSMA} - (1-\beta) \cdot \Delta D^{URLLC,avg}_{USMA})
\end{multline}
where $\omega_{UMSA}$ and $\omega_{MSMA}$ are the weights for considering the effect of USMA and MSMA on the system reward and determine the relative importance of reducing latency or maximizing throughput. In the current work, we assume equal priority for both URLLC and eMBB services, and both assigned a value of 1. Equal weights simplify resource allocation, ensuring a balanced trade-off between low-latency communication and high throughput, while promoting fairness and stability in mixed-service environments where neither service is prioritized over the other. 
$\Delta B^{eMBB,avg}_{MSMA}$ is the change in throughput and $\Delta D^{URLLC,avg}_{USMA}$ is the change in latency due to an agent’s action. This helps the agents internalize the trade-offs. $\beta \in [0,1]$ represents a trade-off factor that adjusts the degree of penalty applied to changes in throughput or latency when agents, take actions. It ensures that while one slice may improve its primary objective (e.g., URLLC reducing latency), the system considers the impact on the other slice (eMBB’s throughput) and balances cooperation between the agents. Essentially, $\beta$ fine-tunes the balance between prioritizing one agent's goal while maintaining fairness across the system. It should be noted that \(\beta = \frac{|\Delta D^{URLLC}|}{|\Delta D^{URLLC}| + |\Delta B^{eMBB}|}\), allows $\beta$ to adaptively prioritize actions based on the relative importance of latency and throughput, facilitating a balanced reward function that reflects the performance goals of both services in the PVDN framework.


As depicted in Fig. \ref{fig:PVDN}, the proposed PVDN learns and adapts MSMA and USMA subtask solutions based on assigned priorities. Using PVDN, the system can reuse previously learned subtask solutions from agents, followed by an adaptation step. PVDN enables the use of retained subtask training data for offline learning, eliminating the need for new environment interactions during adaptation. It is important to note that, compared to VDN, this method reuses and adapts solutions for both URLLC and eMBB slices. Although both agents have their own separate Q-values, they can share knowledge with another agent.

 
\section{Simulation Results}
\label{section:result}

\subsection{Simulation Settings}

In this paper, we consider a scenario of a base station with a cell radius of 125 meters, a bandwidth of 20 MHz, and 13 resource block groups (RBGs). The eNB contains URLLC and eMBB slices with 10 UEs and 5 UEs, respectively which generate traffic by a Poisson distribution. In terms of the learning algorithm parameters, the values of $\alpha$, $\beta$, $\gamma$, and $\epsilon$ are 0.5, 0.5, 0.2, and 0.3, respectively. 

\subsection{Results}

\begin{figure}
    \centering
    \includegraphics[width=0.85\linewidth]{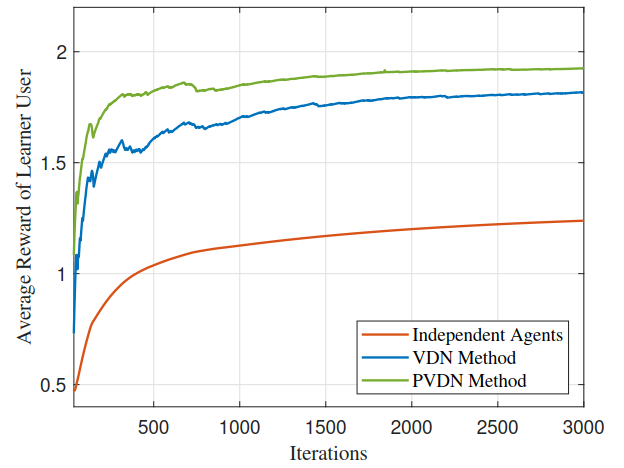}
    \caption{Convergence Comparison of PVDN, VDN, and Independent Methods}
    \label{fig:reward}
    \vspace{-0.7cm}
\end{figure}

In this section, we assess the performance of PVDN wih respect to two baselines, namely, VDN, and independent methods, in meeting the URLLC and eMBB slice requirements. Fig. \ref{fig:reward}  demonstrates the convergence speed of the algorithms. PVDN consistently achieves higher rewards than the other two methods. This improved performance is due to PVDN's ability to prioritize resource allocation between agents, allowing it to learn more efficiently. Additionally, PVDN converges to optimal rewards faster because it dynamically adjusts agent priorities based on system demands, ensuring more effective decision-making throughout the learning process.

\begin{figure*}
\centering
\begin{subfigure}{0.34\textwidth}
    \includegraphics[width=\textwidth]{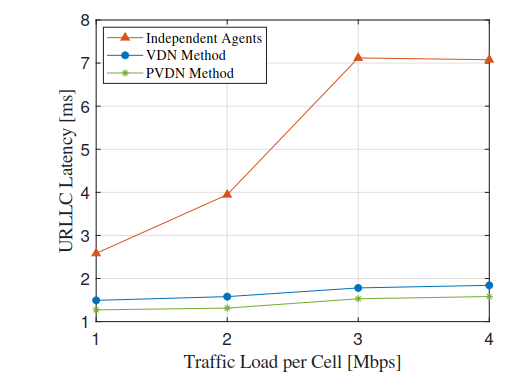}
    \caption{Latency of the URLLC slice under an eMBB traffic load of 1 Mbps}
    \label{fig:URLLC1}
\end{subfigure}
\hfill
\begin{subfigure}{0.32\textwidth}
    \includegraphics[width=\textwidth]{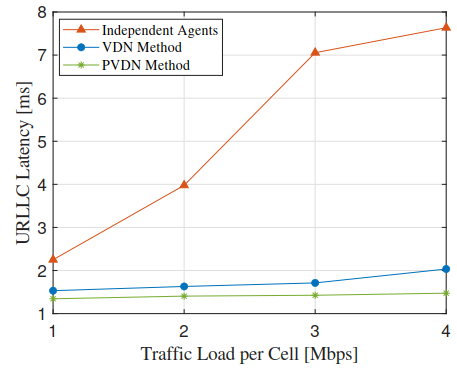}
    \caption{Latency of the URLLC slice under an eMBB traffic load of 2 Mbps}
    \label{fig:URLLC2}
\end{subfigure}
\hfill
\begin{subfigure}{0.32\textwidth}
    \includegraphics[width=\textwidth]{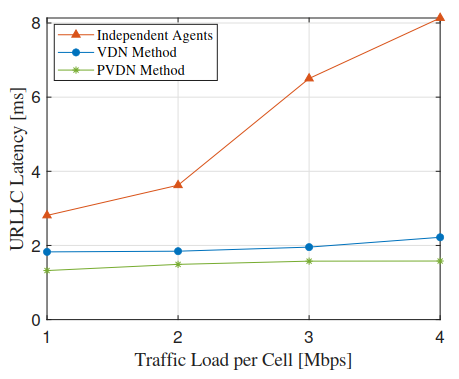}
    \caption{Latency of the URLLC slice under an eMBB traffic load of 3 Mbps}
    \label{fig:URLLC3}
\end{subfigure}
\caption{USMA performance under varying URLLC traffic loads}
\label{Latency_USMA}
\vspace{-15pt}
\end{figure*}

\begin{figure*}
\centering
\begin{subfigure}{0.32\textwidth}
    \includegraphics[width=\textwidth]{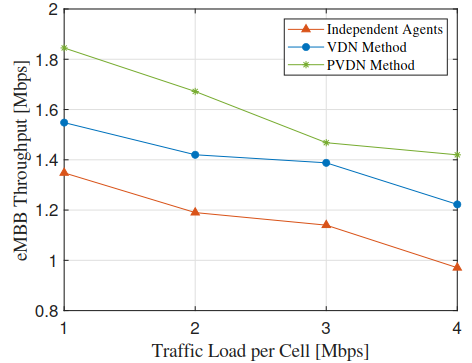}
    \caption{Throughput of the eMBB slice under a URLLC traffic load of 1 Mbps}
    \label{fig:eMBB1}
\end{subfigure}
\hfill
\begin{subfigure}{0.32\textwidth}
    \includegraphics[width=\textwidth]{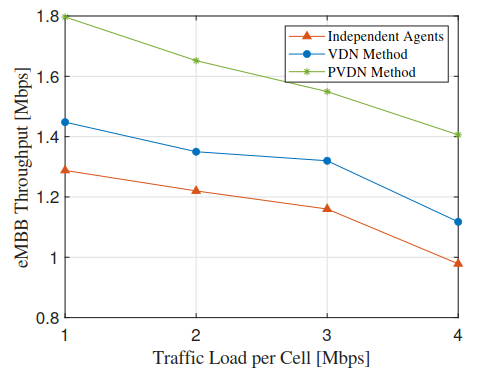}
    \caption{Throughput of the eMBB slice under a URLLC traffic load of 2 Mbps}
    \label{fig:eMBB2}
\end{subfigure}
\hfill
\begin{subfigure}{0.32\textwidth}
    \includegraphics[width=\textwidth]{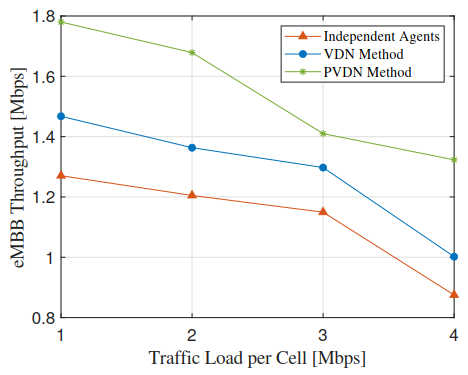}
    \caption{Throughput of the eMBB slice under a URLLC traffic load of 3 Mbps}
    \label{fig:eMBB3}
\end{subfigure}
\caption{MSMA performance under varying eMBB traffic loads}
\label{Throughput_MSMA}
\end{figure*}

Fig. \ref{Latency_USMA} and Fig. \ref{Throughput_MSMA} provide information about the performance of the three algorithms, including independent agents, VDN and PVDN. As depicted in Fig. \ref{Latency_USMA} and Fig. \ref{Throughput_MSMA}, PVDN facilitates effective collaboration, leading to improved resource allocation efficiency, better handling of dynamic network conditions, and enhanced overall performance in 5G network slicing. The PVDN outperforms both the VDN and independent methods in decreasing latency and increasing throughput due to its ability to effectively coordinate resource allocation between USMA and MSMA while respecting the priority of URLLC and eMBB slices. Unlike independent methods, where agents operate in isolation and lack collaboration, PVDN ensures that resource allocation decisions are made with a global view of the system's requirements. The decomposition of the global value function into individual agent contributions allows PVDN to balance the needs of both URLLC (low-latency) and eMBB (high-throughput) slices. By aligning resources with slice-specific priorities, PVDN can minimize delays in the URLLC slice while still maintaining optimal resource allocation for eMBB throughput. This leads to more effective cooperation between agents and improves overall system efficiency.

As depicted in Fig. \ref{fig:URLLC1} to Fig. \ref{fig:URLLC3}, in comparison to VDN, which also enables some level of coordination, PVDN further enhances performance by incorporating prioritization into its learning process. This means that high-priority tasks like reducing latency for URLLC slices are given more attention in resource allocation decisions, without severely impacting eMBB throughput. As a result, PVDN achieves a 35\% reduction in latency compared to the independent method, and 22\% compared to VDN. Similarly, as shown in Fig. \ref{fig:eMBB1} to Fig. \ref{fig:eMBB3}, it improves throughput by 67\% over the independent method and 16\% over VDN. These improvements highlight PVDN's superior capability to adapt and allocate resources efficiently in dynamic network slicing scenarios, making it a more robust solution for real-world applications where latency and throughput are critical performance metrics. 

\section{Conclusion}
\label{section:concl}

In this paper, we 
explored the advantages of utilizing PVDN for resource allocation in an MARL system for network slicing. PVDN offers an efficient approach to managing the two-slices system by breaking down the global value function into individual agent contributions, ensuring efficient resource distribution. Compared to traditional independent methods and VDN, PVDN significantly enhances system performance. These results demonstrate that PVDN's ability to balance the needs of multiple agents within the system makes it an effective solution for resource allocation in complex multi-agent scenarios. 


\section*{Acknowledgment}

This work has been supported by MITACS, Ericsson Canada and partially by NSERC Collaborative Research and Training Experience Program (CREATE) TRAVERSAL under Grant 497981 and Canada Research Chairs program.

\bibliography{references.bib}{}
\bibliographystyle{IEEEtran}

\end{document}